\documentclass[11pt]{article}

\usepackage{amsmath,amssymb,amsfonts}
\usepackage{graphicx}
\usepackage{hyperref}
\usepackage[numbers,sort&compress]{natbib}
\usepackage{booktabs}
\usepackage[left=2cm,right=2cm,top=2cm,bottom=2cm]{geometry}

\usepackage{tikz}
\usetikzlibrary{arrows.meta,positioning}

\hypersetup{
  colorlinks=true,
  linkcolor=blue,
  citecolor=blue,
  urlcolor=blue
}

\newcommand{\dd}{\mathrm{d}}
\newcommand{\ee}{\mathrm{e}}

\newcommand{\RR}{\mathbb{R}}
\newcommand{\HH}{\mathcal{H}}
\newcommand{\DD}{\mathcal{D}}

\tikzset{
  block/.style={draw, rounded corners, align=center, inner sep=6pt, font=\small,
                minimum width=4.3cm, minimum height=1.0cm},
  arrow/.style={-{Stealth[length=2.2mm]}, thick}
}

\title{A Conservative Log-Size Master Equation for Fragmentation PBEs: Jump Transport, Drift--Diffusion Asymptotics, and PSD Inference}

\author{J.\ J.\ Segura\thanks{ORCID: 0009-0004-2742-4353. Corresponding author: \texttt{juan.segura.f@unab.cl}}\\
Universidad Andr\'es Bello
}

\date{}

\begin{document}
\maketitle

\begin{abstract}
Fragmentation population-balance equations (PBEs) describe how particle size distributions (PSDs) evolve under breakage and daughter fragment redistribution. From a standard self-similar fragmentation class we derive an \emph{exact conservative transport equation in log-size} for the \emph{normalized mass fraction}: a state-dependent \emph{pure-jump} master equation (nonlocal internal-coordinate mass transfer). We also give an explicit Gorini--Kossakowski--Sudarshan--Lindblad (GKSL) factorization whose diagonal sector reproduces this master equation, used here as an \emph{optional} structure-preserving operator representation and constrained parameterization for inverse modeling (rather than a computational necessity).

In a controlled small-jump regime, the nonlocal jump transport reduces to a drift--diffusion (Fokker--Planck) operator in log-size space. Under detailed-balance conditions this operator admits the standard symmetrization to a self-adjoint Schr\"odinger-type spectral problem, enabling compact parametric hypothesis classes for PSD shapes. We then present two inverse routes: (i) time-resolved parametric fitting of transport/spectral parameters, and (ii) a regularized steady-state inversion that reconstructs an effective potential from a measured steady PSD.

To address practical validation, we include numerical benchmarks: forward simulation of the jump transport model (CTMC discretization) and its drift--diffusion reduction, quantitative discrepancy metrics, and inverse parameter recovery on an Airy half-line synthetic benchmark under controlled multiplicative noise.
\end{abstract}


\section{Introduction}
Fragmentation and comminution evolve an ensemble of particles through an \emph{internal coordinate} (size, mass, or a proxy), and the particle size distribution (PSD) is a central state variable for design, control, and optimization of grinding and classification circuits. In pure-breakage PBEs, material is removed from a parent size at a selection rate and redistributed nonlocally to smaller sizes through a daughter kernel. This combination is naturally interpreted as \emph{transport in internal-coordinate space}: a conservative gain--loss (jump) generator in which nonnegativity and mass conservation are structural.

Classical fragmentation equations and PBEs have a long history, including self-similar solutions for homogeneous kernels and widely used empirical PSD families such as Rosin--Rammler/Weibull, log-normal, log-logistic, and Swebrec-type forms \citep{Rosin1933,Kapur1972,OBrien2005,Ziff1985,Filippov1961}. The objective of this article is to organize a \emph{transport-first} formulation in log-size space that is explicit about (i) conservative normalization, (ii) the relation between nonlocal jump transport and drift--diffusion reductions, and (iii) practically implementable inverse routes.

A secondary objective is to show that a GKSL/Lindblad form can be used as an \emph{optional} operator factorization of the classical Markov generator. We do \emph{not} claim GKSL is required to simulate PBEs---standard CTMC and finite-volume discretizations already preserve positivity and conservation. Rather, the GKSL factorization is useful as (i) a constrained parameterization for inverse problems (nonnegativity is built in through squared amplitudes) and (ii) a compact semigroup/spectral language that connects jump transport, diffusion limits, and reduced spectral models.

Potential application domains include mineral comminution and classification circuits (where PSD evolution is central for control), pharmaceutical milling (PSD-driven product performance), and other particulate fragmentation settings where transport in internal-coordinate space is the natural modeling language.

\paragraph{What is new (and what is not).}
The log transform for homogeneous/self-similar fragmentation is classical (e.g.\ early work such as \citep{Filippov1961}). The contributions here are:
\begin{itemize}
\item An explicit conservative \emph{mass-fraction} normalization in log-size space that yields a Markov jump generator (Sec.~\ref{sec:logjump}).
\item An explicit GKSL/Lindblad \emph{factorization} of that classical generator (Sec.~\ref{sec:gksl}), used as an optional structure-preserving operator representation and constrained inverse-model parameterization.
\item A controlled small-jump reduction to drift--diffusion (Sec.~\ref{sec:diffusion}) and a practical spectral parametrization of PSD shapes (Sec.~\ref{sec:dictionary}).
\item Numerical validation (Sec.~\ref{sec:validation}): forward jump vs.\ drift--diffusion comparison and inverse recovery under noise on a well-posed synthetic benchmark.
\end{itemize}

\begin{figure}[t]
\centering
\resizebox{\linewidth}{!}{%
\begin{tikzpicture}[node distance=8mm and 10mm]
\tikzset{
  block/.style={
    draw, rounded corners,
    text width=4.0cm,
    align=center,
    inner sep=4pt,
    font=\footnotesize,
    minimum height=1.0cm
  },
  arrow/.style={-{Stealth[length=2.2mm]}, thick}
}

\node[block] (pbe) {PBE in size space $x$\\(selection + daughter redistribution)};
\node[block, right=of pbe] (log) {Log-size transport $\xi=\ln(x/x_0)$\\conservative mass-fraction generator};
\node[block, below=of log] (jump) {Exact jump transport\\(master equation)\\Eq.~\eqref{eq:master}};
\node[block, below=of jump] (fp) {Small-jump reduction\\drift--diffusion\\Eq.~\eqref{eq:FP}};
\node[block, left=of fp] (gksl) {Optional GKSL factorization\\constrained inverse parametrization};
\node[block, right=of fp] (inv) {Inverse routes + validation\\Secs.~\ref{sec:inverse}, \ref{sec:validation}};

\draw[arrow] (pbe) -- (log);
\draw[arrow] (log) -- (jump);
\draw[arrow] (jump) -- (fp);
\draw[arrow] (fp) -- (inv);
\draw[arrow] (jump) -- (gksl);
\end{tikzpicture}%
}
\caption{Modeling framework: a transport-theoretic approach to fragmentation PBEs in log-size space. The GKSL/Lindblad form is optional and used primarily as a structure-preserving factorization and constrained parameterization for inverse modeling; forward simulation is performed directly on the classical jump/FP transport models and validated numerically.}
\label{fig:pipeline}
\end{figure}
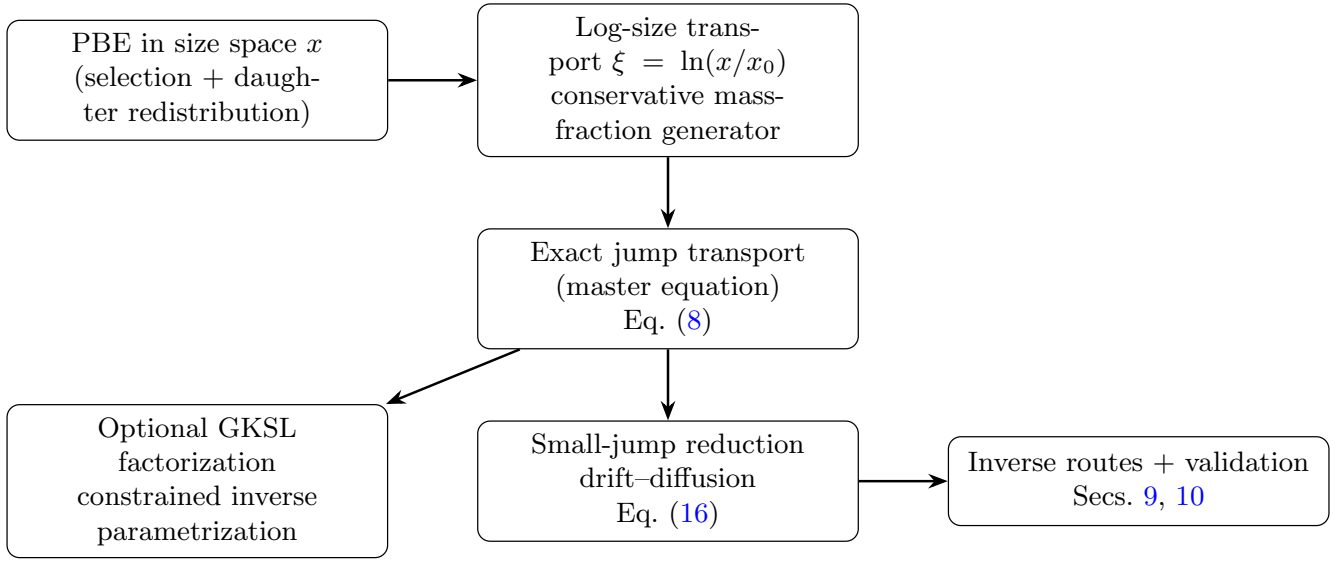

\section{Related work and context}
Log-space formulations of self-similar fragmentation and conservative gain--loss generators are classical in fragmentation theory, and we do not claim novelty for the log transform itself (see, e.g., \citep{Filippov1961}). Our emphasis is instead on (i) the conservative \emph{mass-fraction} normalization in log-space that yields a Markov generator with transparent positivity and conservation, (ii) an explicit GKSL/Lindblad factorization of that classical generator used as an optional constrained parameterization for inverse modeling, and (iii) validated inverse workflows demonstrated numerically.

Inverse identification of fragmentation parameters/kernels from data has recently received renewed attention; see, e.g., \citep{Doumic2024} for kernel recovery from short-time behavior and data-driven discovery approaches such as \citep{Ali2025,Tiong2025}. Operator methods for classical stochastic processes also have long precedent, including second-quantization and coherent-state/path-integral formalisms \citep{Doi1976,Peliti1985}. Here, GKSL/Lindblad notation is used neither as a microscopic quantum claim nor as a replacement for standard PBE numerics, but as a convenient \emph{factorization language} for conservative Markov generators and for connecting jump transport, diffusion limits, and reduced spectral parametrizations.

\section{Nomenclature}
\begin{tabular}{@{}ll@{}}
\toprule
Symbol & Meaning \\ \midrule
$x$ & size-like coordinate (taken proportional to conserved mass here) \\
$n(x,t)$ & number density in $x$ \\
$m(x,t)=x\,n(x,t)$ & mass density in $x$ \\
$M=\int_0^\infty m(x,t)\dd x$ & total mass (conserved) \\
$\xi=\ln(x/x_0)$ & log-size coordinate \\
$p(\xi,t)$ & normalized mass-fraction density in $\xi$ \\
$S(x)$ & selection/breakage rate \\
$b_n(x,y)$, $b_m(x,y)$ & daughter kernels (number / mass-weighted) \\
$\lambda(\xi)$ & jump rate in $\xi$ \\
$K(u)$ & jump-length kernel in $\xi$ \\
\bottomrule
\end{tabular}

\section{Fragmentation transport model and the log-size jump process}
\label{sec:logjump}

\subsection{Number density, mass density, and the mass-weighted kernel}
Let $x>0$ denote a size-like variable. For definiteness, take $x$ proportional to particle mass (or any conserved extensive variable). A standard fragmentation equation for the \emph{number density} $n(x,t)$ is
\begin{equation}
\partial_t n(x,t)=-S(x)n(x,t)+\int_x^\infty b_n(x,y)\,S(y)n(y,t)\,\dd y,
\label{eq:PBE-number}
\end{equation}
where $S(x)$ is the selection (breakage) rate and $b_n(x,y)$ is the daughter \emph{number} density. Mass conservation requires
\begin{equation}
\int_0^y x\,b_n(x,y)\,\dd x = y.
\label{eq:masscons-number}
\end{equation}

Define the \emph{mass density} $m(x,t)=x\,n(x,t)$. Then \eqref{eq:PBE-number} implies
\begin{equation}
\partial_t m(x,t)
=-S(x)m(x,t)+\int_x^\infty b_m(x,y)\,S(y)m(y,t)\,\dd y,
\label{eq:PBE-mass}
\end{equation}
where the \emph{mass-weighted} daughter kernel is
\begin{equation}
b_m(x,y)=\frac{x}{y}\,b_n(x,y),
\qquad
\int_0^y b_m(x,y)\dd x=1.
\label{eq:bm-def}
\end{equation}
Thus $b_m$ redistributes the parent's mass over daughter sizes; it is the natural kernel for mass-fraction transport. Let $M=\int_0^\infty m(x,t)\dd x$ be the total mass; then $\dd M/\dd t=0$ under \eqref{eq:PBE-mass} and \eqref{eq:bm-def}.

\subsection{Self-similar (homogeneous) kernel}
Assume the widely used self-similar structure
\begin{equation}
S(x)=k\,x^\alpha,
\qquad
b_m(x,y)=\frac{1}{y}\,B\!\left(\frac{x}{y}\right),
\qquad 0<x<y,
\label{eq:selfsimilar}
\end{equation}
where $k>0$, $\alpha\in\RR$, and $B(z)$ is supported on $z\in(0,1)$ with normalization
\begin{equation}
\int_0^1 B(z)\,\dd z = 1.
\label{eq:B-norm}
\end{equation}

\subsection{Log-size transform and an exact conservative master equation}
Let $\xi=\ln(x/x_0)$ so that $x=x_0\ee^\xi$ and $\dd x = x\,\dd\xi$.
Define the normalized \emph{mass fraction density in log-space}
\begin{equation}
p(\xi,t)=\frac{m(x,t)}{M}\,x,
\qquad
\int_{-\infty}^{\infty}p(\xi,t)\,\dd\xi=1.
\label{eq:pdef}
\end{equation}
Equivalently, $p(\xi,t)\dd\xi$ is the mass fraction in the log-bin $(\xi,\xi+\dd\xi)$.

A direct change of variables in \eqref{eq:PBE-mass}--\eqref{eq:selfsimilar} yields the \emph{exact} gain--loss form
\begin{equation}
\partial_t p(\xi,t)
=
-\lambda(\xi)\,p(\xi,t)
+\int_0^\infty \lambda(\xi+u)\,K(u)\,p(\xi+u,t)\,\dd u,
\label{eq:master}
\end{equation}
where
\begin{equation}
\lambda(\xi)=S(x_0\ee^\xi)=k\,x_0^\alpha\,\ee^{\alpha\xi},
\qquad
K(u)=\ee^{-u}\,B(\ee^{-u}),\quad u\ge 0.
\label{eq:lambdaK}
\end{equation}
Using \eqref{eq:B-norm} one checks
\begin{equation}
\int_0^\infty K(u)\,\dd u=1,
\label{eq:Knorm}
\end{equation}
so \eqref{eq:master} preserves $\int p\,\dd\xi$ exactly. Equation \eqref{eq:master} is therefore a conservative, state-dependent \emph{one-sided jump transport} equation in log-size space.

\section{Optional operator factorization: a GKSL/Lindblad dilation of the jump generator}
\label{sec:gksl}

\paragraph{Role of the GKSL/Lindblad form.}
The classical master equation \eqref{eq:master} is already a conservative Markov transport model and can be simulated directly with CTMC or finite-volume discretizations. We therefore do \emph{not} present GKSL/Lindblad notation as a computational requirement. Its role here is as an \emph{optional factorization} that (i) provides a constrained parameterization for inverse modeling (nonnegativity built in through squared amplitudes) and (ii) connects jump transport, diffusion limits, and spectral reductions within a semigroup framework.

\subsection{A GKSL construction reproducing the master equation on the diagonal}
Consider $\HH=L^2(\RR)$ with generalized position basis $\{|\xi\rangle\}$. Let $\rho(t)$ be a positive trace-class operator and define the diagonal
\begin{equation}
p(\xi,t)=\langle\xi|\rho(t)|\xi\rangle.
\label{eq:diag}
\end{equation}
(Here $\langle\xi|\rho(t)|\xi\rangle$ denotes the diagonal element of the operator $\rho(t)$ in the $\xi$ representation, i.e.\ $p(\xi,t)=\rho(\xi,\xi,t)$.)

Define a nonlocal jump operator $L$ by its kernel
\begin{equation}
(L\psi)(\xi)
=\int_{\xi}^{\infty}\sqrt{\lambda(\eta)}\,\sqrt{K(\eta-\xi)}\,\psi(\eta)\,\dd\eta.
\label{eq:Ldef}
\end{equation}
Now consider the GKSL (Lindblad) equation
\begin{equation}
\frac{\dd\rho}{\dd t}
=
-i[H_0,\rho]
+\DD[\rho],
\qquad
\DD[\rho]=L\rho L^\dagger-\tfrac{1}{2}\{L^\dagger L,\rho\},
\label{eq:GKSL}
\end{equation}
with any Hermitian $H_0$ (set $H_0=0$ for a purely dissipative factorization).

\paragraph{Claim (Diagonal reproduction).}
If $\rho(t)$ is diagonal in the $|\xi\rangle$ basis, i.e.\ $\rho(\eta,\eta')=p(\eta,t)\delta(\eta-\eta')$, then $p(\xi,t)$ satisfies the classical master equation \eqref{eq:master}.
\emph{Sketch.} Under diagonality,
$\langle\xi|L\rho L^\dagger|\xi\rangle
=\int_0^\infty \lambda(\xi+u)K(u)p(\xi+u,t)\dd u$,
and $(L^\dagger L)(\xi,\xi)=\lambda(\xi)$ using \eqref{eq:Knorm}. \hfill$\square$

\paragraph{Practical payoff for inverse modeling.}
A common failure mode in kernel learning is violating nonnegativity or normalization. The factorization in \eqref{eq:Ldef} builds these constraints in through the square-root structure $\sqrt{\lambda}\sqrt{K}$, providing a convenient constrained parameterization for inverse problems without requiring ad hoc projection steps.

\section{Diffusive (small-jump) limit: drift--diffusion transport in log-size}
\label{sec:diffusion}

When $K(u)$ concentrates near $u=0$, \eqref{eq:master} admits a controlled Kramers--Moyal expansion. Define moments
\begin{equation}
m_n=\int_0^\infty u^n K(u)\,\dd u,\qquad m_0=1,
\label{eq:moments}
\end{equation}
and assume $m_1,m_2<\infty$ and higher moments are sufficiently small.
Expanding $\lambda(\xi+u)p(\xi+u,t)$ and using $m_0=1$ cancels the loss exactly, yielding at second order
\begin{equation}
\partial_t p(\xi,t)
=
m_1\,\partial_\xi\big(\lambda p\big)
+\frac{m_2}{2}\,\partial_{\xi\xi}\big(\lambda p\big)
+\mathcal{O}(m_3).
\label{eq:FP-raw}
\end{equation}
This can be written in conservative Fokker--Planck form
\begin{equation}
\partial_t p
=-\partial_\xi\big(A(\xi)p\big)+\partial_{\xi\xi}\big(D(\xi)p\big),
\qquad
A(\xi)=-m_1\lambda(\xi),\quad D(\xi)=\frac{m_2}{2}\lambda(\xi).
\label{eq:FP}
\end{equation}

\section{Spectral form (detailed balance case) and PSD parametrizations}
\label{sec:spectral}

If the FP operator admits a zero-current steady state (detailed balance), there exists $\Phi(\xi)$ such that
\begin{equation}
A(\xi)=2D(\xi)\,\Phi'(\xi).
\label{eq:DB}
\end{equation}
Then the similarity transform
\begin{equation}
p(\xi,t)=\ee^{-\Phi(\xi)}\,\psi(\xi,t)
\label{eq:sim}
\end{equation}
removes the first-derivative term and yields an imaginary-time Schr\"odinger form (standard; see \citep{Risken1989})
\begin{equation}
\partial_t\psi
=
D(\xi)\,\partial_{\xi\xi}\psi
-U_{\mathrm{eff}}(\xi)\,\psi.
\label{eq:imagSchr}
\end{equation}
In this detailed-balance case, the stationary density satisfies
\begin{equation}
p_{\mathrm{ss}}(\xi)\propto \ee^{-\Phi(\xi)}=\phi_0(\xi)^2,
\label{eq:ss-square}
\end{equation}
where $\phi_0$ is the ground state of the symmetrized operator. Non-self-adjoint/open steady states are briefly discussed in Appendix~\ref{app:biorth}.

\section{Parametric hypothesis classes for PSD shapes}
\label{sec:dictionary}

The mapping from an empirical PSD shape to a unique generator is generally non-identifiable from a single steady snapshot. The purpose of this section is therefore \emph{not} unique identification, but a compact set of explicit low-dimensional hypothesis classes for inverse fitting (especially with time-resolved data).

\begin{table}[t]
\centering
\caption{Representative PSD shapes in $\xi$ and explicit potential families (detailed-balance / self-adjoint reduction). These serve as parametric hypothesis classes for inverse fitting, not unique identifications.}
\label{tab:dict}
\begin{tabular}{@{}lll@{}}
\toprule
PSD shape in $\xi$ & Effective potential $V(\xi)$ (up to scaling) & Typical PSD in $x$ \\ \midrule
Gaussian & $V(\xi)=\tfrac{1}{2}\omega^2(\xi-\xi_0)^2$ & Log-normal \\
$\mathrm{sech}^2$ core & $V(\xi)=-V_0\,\mathrm{sech}^2(a(\xi-\xi_0))$ (P\"oschl--Teller) & Log-logistic \\
Exponential-in-$\ee^{a\xi}$ tail & $V(\xi)=V_0\left(\ee^{-2a\xi}-2\ee^{-a\xi}\right)$ (Morse-type) & Weibull/Rosin--Rammler-like \\
Power-law regime & $V(\xi)=g/(\xi-\xi_0)^2$ (scale-invariant) & Power-law tails \\
Multi-shoulder/multi-slope & multi-well or low-mode mixture & Swebrec-like flexibility \\ \bottomrule
\end{tabular}
\end{table}

\section{Inverse modeling routes}
\label{sec:inverse}

\subsection{Route I: time-resolved parametric fitting}
Choose a parametric family for the transport generator, e.g.\ via $\Phi(\xi;\theta)$ (detailed balance FP) or via a constrained kernel family $K(u;\theta)$ for the jump model. Fit $\theta$ to time-resolved PSD snapshots $\tilde p(\xi,t_j)$ by minimizing a discrepancy, e.g.\ weighted least squares or a discretized KL divergence, subject to normalization/positivity constraints. The optional factorization in Sec.~\ref{sec:gksl} provides a convenient constrained parameterization for jump kernels and rates.

\subsection{Route II: steady inversion with explicit regularization}
Assume a steady shape is observed and the detailed-balance reduction is adequate so $p_{\mathrm{ss}}(\xi)\propto \phi_0(\xi)^2$ with $\phi_0>0$. For a constant-mass Schr\"odinger form
\[
-\frac{\hbar^2}{2\mu}\phi_0''(\xi)+V(\xi)\phi_0(\xi)=E_0\phi_0(\xi),
\]
In the fragmentation context, the ratio $\hbar^2/2\mu$ plays the role of an effective diffusivity scale in the symmetrized reduction, and may be identified with $D(\xi)$ in Eq.~\eqref{eq:FP} (or with a constant $D_0$ after a Lamperti/Liouville transform); in turn $D(\xi)=(m_2/2)\lambda(\xi)$ relates it back to the jump kernel moments and rate via Eqs.~\eqref{eq:moments} and \eqref{eq:FP}.
one can reconstruct
\begin{equation}
V(\xi)=E_0+\frac{\hbar^2}{2\mu}\frac{\phi_0''(\xi)}{\phi_0(\xi)}
=E_0+\frac{\hbar^2}{2\mu}\left[\frac{p_{\mathrm{ss}}''}{2p_{\mathrm{ss}}}-\frac{(p_{\mathrm{ss}}')^2}{4p_{\mathrm{ss}}^2}\right].
\label{eq:directV}
\end{equation}
Because derivatives amplify noise, we recommend smoothing $p_{\mathrm{ss}}$ by a curvature-penalized (Tikhonov) estimate
\[
p^\star=\arg\min_p \ \|p-\tilde p\|_2^2+\lambda\|p''\|_2^2,
\]
with $\lambda$ chosen by an L-curve or generalized cross-validation; Sec.~\ref{sec:validation} reports noise-sweep behavior for a synthetic benchmark.

\section{Numerical validation: forward prediction, reduction accuracy, and inverse recovery}
\label{sec:validation}

This section addresses practical validation requested by transport-phenomena reviewers: (i) forward simulation of the classical jump transport model \eqref{eq:master}, (ii) quantitative comparison against the drift--diffusion reduction \eqref{eq:FP}, and (iii) inverse parameter recovery from noisy synthetic PSDs.

\subsection{Forward benchmark: jump transport (CTMC discretization)}
\begin{figure}[t]
\centering
\includegraphics[width=0.85\linewidth]{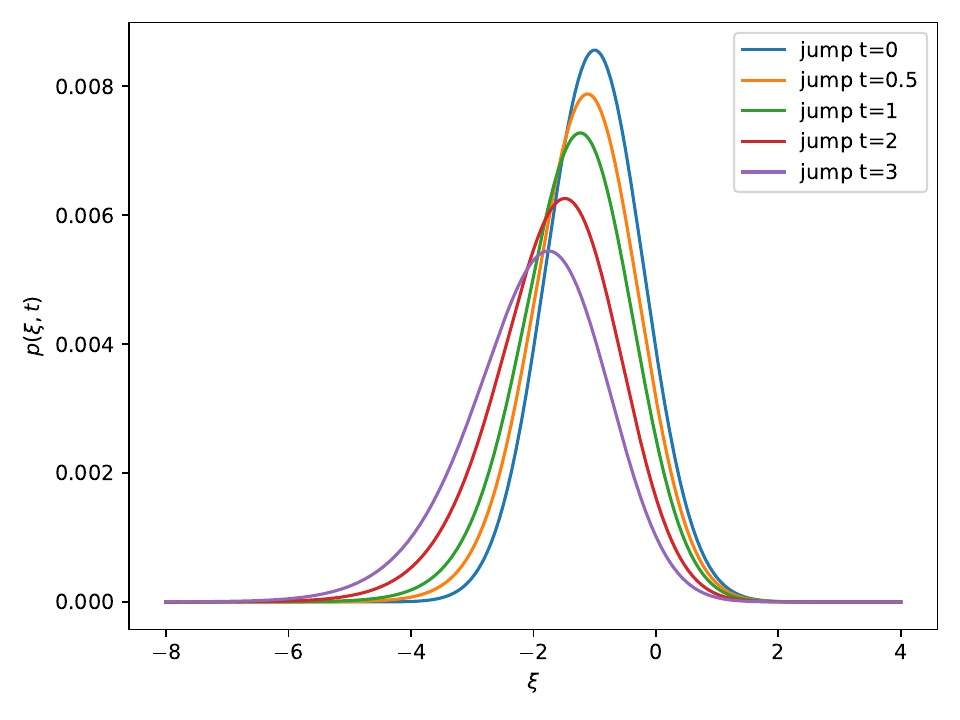}
\caption{Forward simulation of the log-size jump transport model (CTMC discretization of Eq.~\eqref{eq:master}) at representative times. The solution remains nonnegative and normalized by construction.}
\label{fig:forward_jump}
\end{figure}

We discretize $\xi$ on a uniform grid and evolve bin masses with a conservative continuous-time Markov chain (CTMC) generator $Q$ (column sums zero; off-diagonals nonnegative). This guarantees nonnegativity and conservation at the discrete level. Figure~\ref{fig:forward_jump} shows representative forward snapshots for the jump model.

\subsection{Forward benchmark: drift--diffusion reduction}
\begin{figure}[t]
\centering
\includegraphics[width=0.85\linewidth]{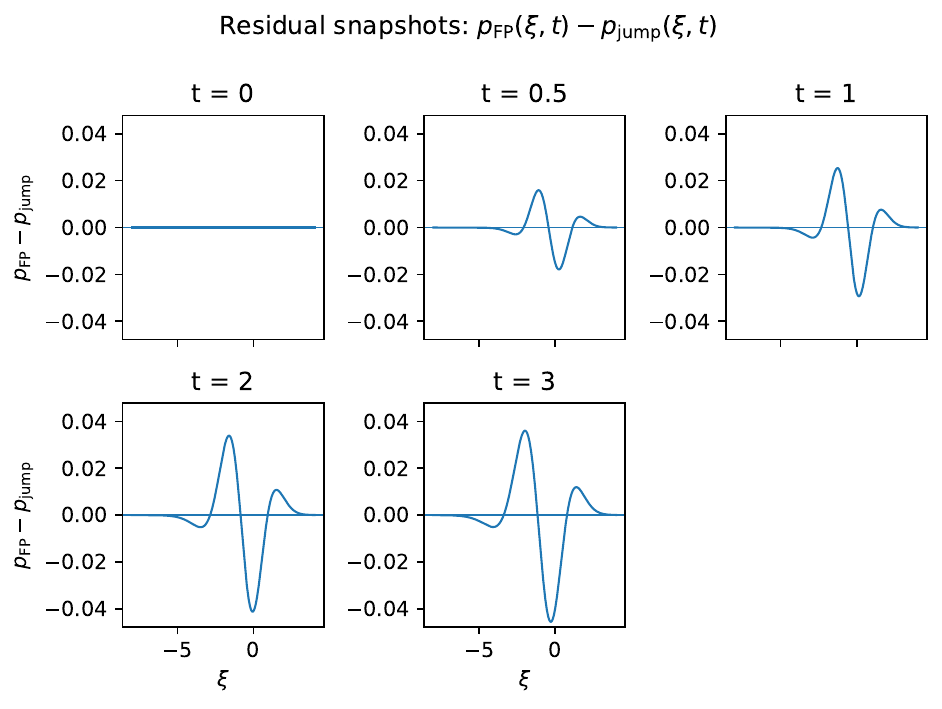}
\caption{Drift--diffusion reduction residual relative to the jump transport solution at the same times as Fig.~\ref{fig:forward_jump}: $p_{\mathrm{FP}}(\xi,t)-p_{\mathrm{jump}}(\xi,t)$. This highlights where the Fokker--Planck reduction deviates from the nonlocal jump transport model.}
\label{fig:forward_fp}
\end{figure}

We solve the FP reduction \eqref{eq:FP} on the same $\xi$ grid using a conservative flux discretization and compare against the jump solution. Figure~\ref{fig:forward_fp} shows representative FP snapshots at the same times as Fig.~\ref{fig:forward_jump}.

\subsection{Reduction accuracy: jump vs.\ drift--diffusion discrepancy metrics}
\begin{figure}[t]
\centering
\includegraphics[width=0.85\linewidth]{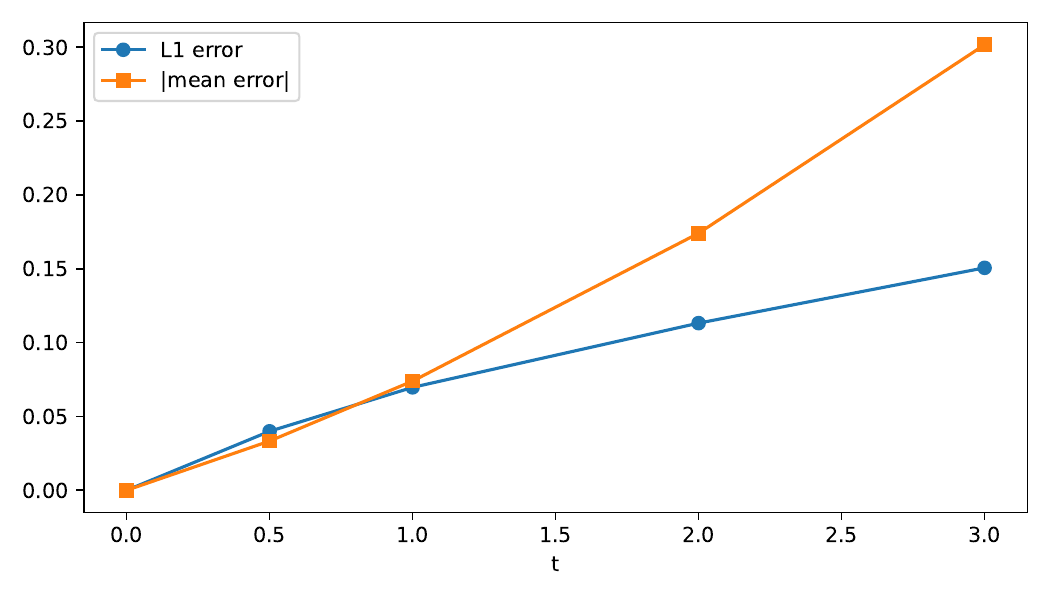}
\caption{Jump vs.\ drift--diffusion validation in log-size space. Shown are $\|p_{\mathrm{jump}}-p_{\mathrm{FP}}\|_1$ and $|\langle\xi\rangle_{\mathrm{jump}}-\langle\xi\rangle_{\mathrm{FP}}|$ as functions of time, illustrating the regime of agreement of the Fokker--Planck reduction.}
\label{fig:jump_vs_fp}
\end{figure}

To quantify the regime of agreement of the FP reduction, we compute the $L^1$ discrepancy in $\xi$,
$\|p_{\mathrm{jump}}-p_{\mathrm{FP}}\|_1$,
and a moment-level discrepancy in the mean log-size $|\langle\xi\rangle_{\mathrm{jump}}-\langle\xi\rangle_{\mathrm{FP}}|$.
Figure~\ref{fig:jump_vs_fp} summarizes these metrics over time for a representative small-jump configuration.

\begin{figure}[t]
\centering
\includegraphics[width=0.75\linewidth]{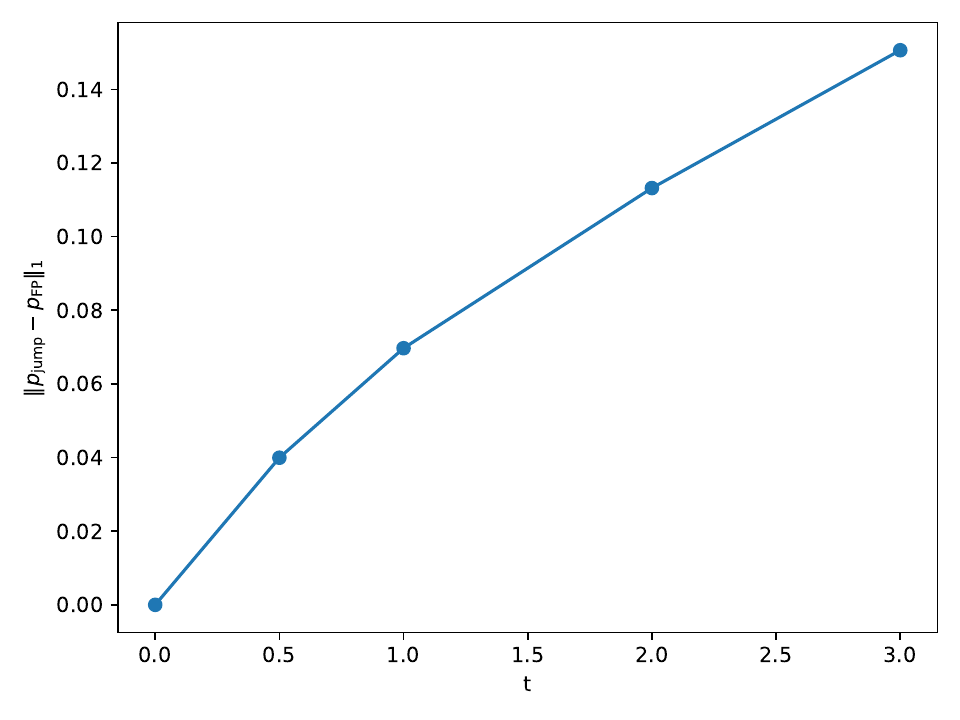}
\caption{Supplementary jump vs.\ Fokker--Planck comparison: $L^1$ discrepancy norm $\|p_{\mathrm{jump}}-p_{\mathrm{FP}}\|_1$ versus time (single-metric view).}
\label{fig:jump_vs_fp_supp}
\end{figure}

\paragraph{Supplementary single-metric view.}
For completeness, Fig.~\ref{fig:jump_vs_fp_supp} provides a single-metric discrepancy view (useful as a compact supplement).

\subsection{Inverse recovery benchmark: Airy half-line synthetic test under noise}
\begin{figure}[t]
\centering
\includegraphics[width=0.75\linewidth]{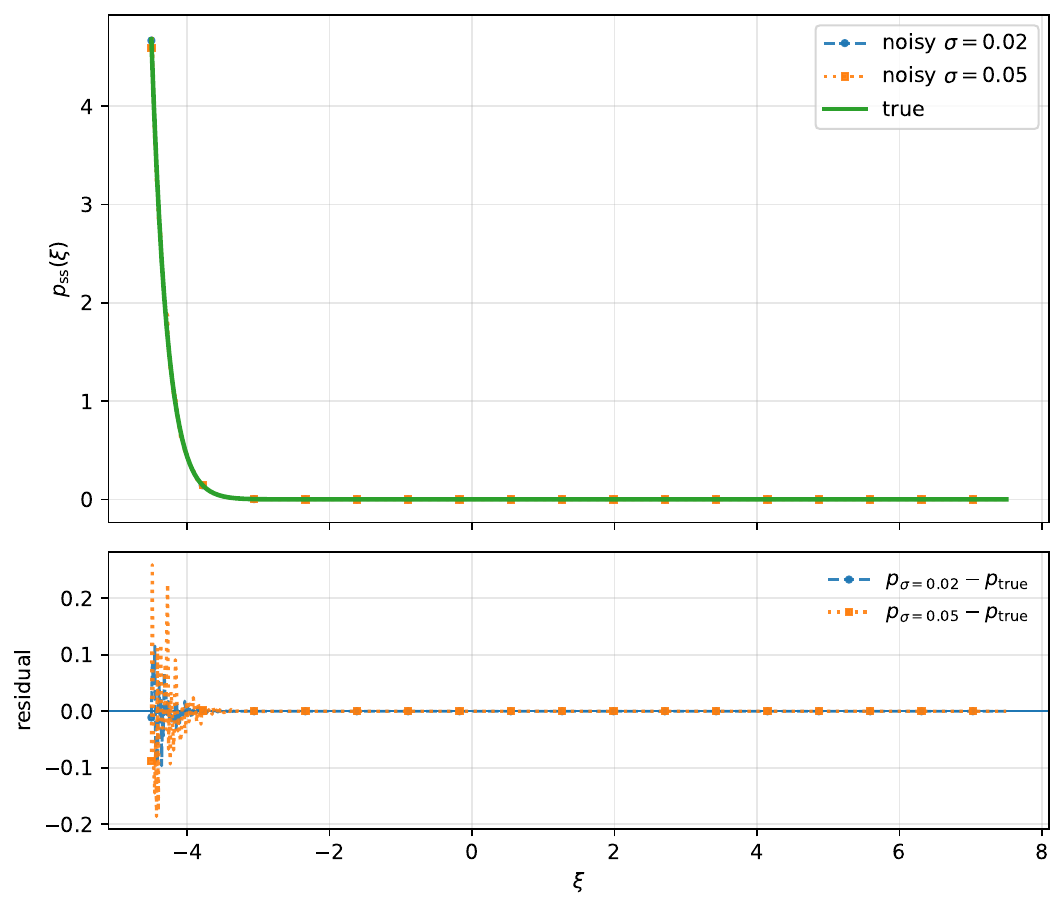}
\caption{Synthetic steady PSD in log-size space for the Airy half-line benchmark.
\textbf{Top:} clean signal and two noisy realizations (multiplicative noise at $\sigma=0.02,0.05$).
\textbf{Bottom:} residuals (noisy $-$ true) showing noise structure. Noisy PSDs are renormalized before inverse fitting.}
\label{fig:airy_noisy}
\end{figure}

\begin{figure}[t]
\centering
\includegraphics[width=0.88\linewidth]{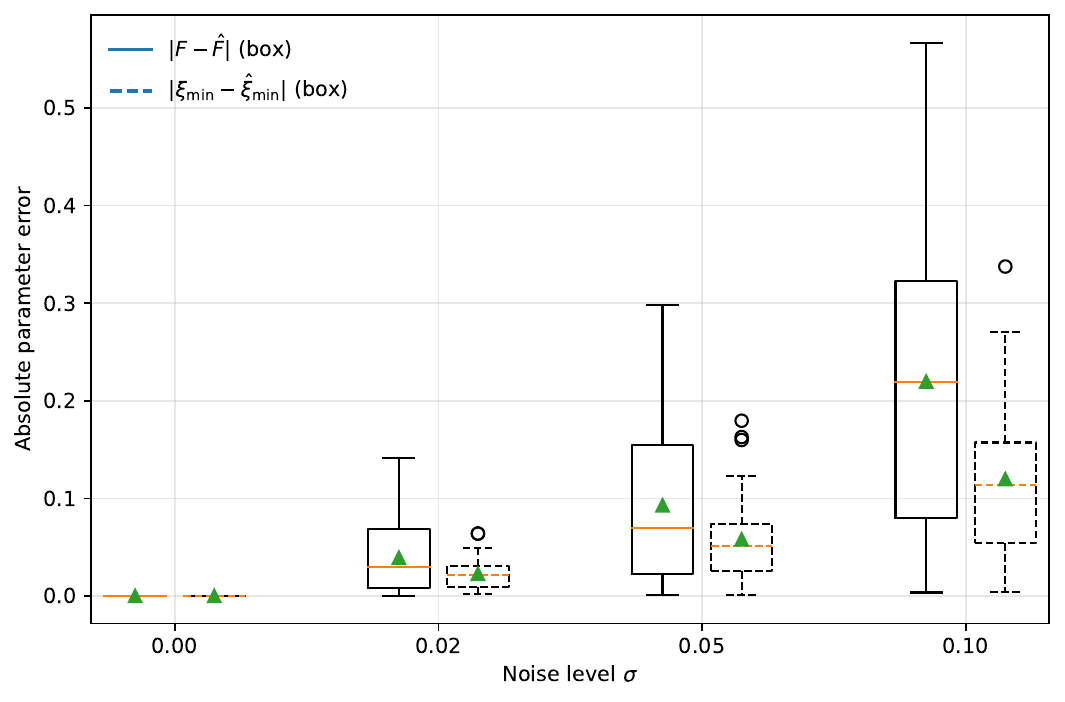}
\caption{Inverse recovery benchmark (Airy half-line synthetic test): distribution of absolute parameter errors versus multiplicative noise level $\sigma$ across repeated noise realizations (box: median and interquartile range; whiskers: 1.5 IQR; markers: means).}
\label{fig:inverse_noise}
\end{figure}

We validate inverse recovery on a well-posed half-line Airy model (Sec.~\ref{sec:example}). Synthetic steady PSDs are generated, contaminated with multiplicative noise, renormalized, and then used to fit parameters. Figure~\ref{fig:airy_noisy} shows typical noisy realizations. Figure~\ref{fig:inverse_noise} reports absolute parameter errors versus noise level.

\paragraph{Inverse identifiability (qualitative).}
The approximately linear growth of absolute errors with noise level in Fig.~\ref{fig:inverse_noise} is consistent with the well-posed nature of this controlled inverse benchmark. The two parameters $(F,\xi_{\min})$ exhibit moderate correlation but remain identifiable in practice: $F$ primarily controls the tail steepness while $\xi_{\min}$ shifts the effective support/peak location.

\subsection{Implementation details (discretization, time stepping, regularization)}
For Figs.~\ref{fig:forward_jump}--\ref{fig:jump_vs_fp_supp}, we use a uniform log-size grid $\xi\in[-8,4]$ with $N_\xi=700$ points and output times $t\in\{0,0.5,1,2,3\}$. The jump benchmark uses a one-sided exponential jump kernel $K(u)=\beta\,\ee^{-\beta u}$ with $\beta=3$ and a constant jump rate $\lambda(\xi)\equiv\lambda_0=1$ (a simple configuration chosen to isolate jump vs.\ drift--diffusion behavior). Time stepping for both the jump transport and the FP reduction uses a stable explicit update with $\Delta t=5\times10^{-4}$; after each step we enforce nonnegativity by clipping at zero and renormalize to $\int p(\xi,t)\dd\xi=1$.

For Fig.~\ref{fig:airy_noisy} (Airy half-line steady benchmark), we use $\xi_{\min}=-4.5$, $\xi\in[\xi_{\min},\xi_{\min}+12]$, and $N_\xi=1000$, with the ground-state parameter $F=1.3$. Multiplicative (lognormal) noise is applied and the noisy PSD is renormalized before fitting. For Fig.~\ref{fig:inverse_noise} we use $R=40$ independent noise realizations per noise level $\sigma\in\{0,0.02,0.05,0.10\}$ and report absolute parameter errors. For Route~II (Sec.~\ref{sec:inverse}), the regularization parameter in $\|p-\tilde p\|_2^2+\lambda\|p''\|_2^2$ is selected automatically by an L-curve or generalized cross-validation; for the Airy benchmark at $N_\xi=1000$ and $\sigma=0.05$, GCV typically selects $\lambda$ on the order of $10^{-8}$ (grid-dependent through $\Delta\xi$).

\paragraph{Computational notes.}
The one-sided jump structure yields a sparse/triangular gain operator in discretized form, and CTMC evolution can be performed stably with implicit stepping, uniformization, or matrix-free exponential methods. The FP reduction can be evolved with conservative flux discretizations. The GKSL factorization in Sec.~\ref{sec:gksl} is not needed for forward evolution and is used primarily for constrained inverse parameterizations.

\section{Worked synthetic example: a well-posed Airy-type half-line model}
\label{sec:example}

The Airy ansatz must be posed in a square-integrable setting: on $\RR$, $\mathrm{Ai}$ oscillates as $\xi\to-\infty$. A clean fix is to work on a half-line with cutoff $\xi_{\min}$ (minimum resolved size) and a confining potential.

\subsection{Forward model}
Let $\xi\in[\xi_{\min},\infty)$ and consider the self-adjoint Hamiltonian
\begin{equation}
H=-\frac{\hbar^2}{2\mu}\frac{\dd^2}{\dd\xi^2}+F(\xi-\xi_{\min}),
\qquad F>0,
\label{eq:airy-halfline}
\end{equation}
with Dirichlet boundary $\phi(\xi_{\min})=0$.
Eigenfunctions are shifted Airy functions
\[
\phi_n(\xi)=\mathcal{N}_n\,\mathrm{Ai}\!\Big( a(\xi-\xi_{\min})-a_n\Big),
\quad
a=\left(\frac{2\mu F}{\hbar^2}\right)^{1/3},
\]
where $a_n>0$ are the zeros of $\mathrm{Ai}(-a_n)=0$. The ground state is positive for $\xi>\xi_{\min}$ and yields a valid PSD:
\begin{equation}
p(\xi)=\phi_0(\xi)^2,\qquad \int_{\xi_{\min}}^\infty p(\xi)\,\dd\xi=1.
\label{eq:airy-psd}
\end{equation}

\subsection{Synthetic data generation and inverse fit}
Synthetic steady PSDs are generated from \eqref{eq:airy-psd}, broadened (optional) and contaminated with multiplicative noise, then renormalized. Parameters $(F,\xi_{\min})$ are fitted by minimizing a discrepancy between predicted and observed binned densities; the resulting noise sweep is reported in Sec.~\ref{sec:validation}.

\section{Limitations and validation targets}
The scope is restricted to pure breakage (no aggregation, growth, classification), self-similar kernels, and diffusion reduction only when jump lengths are small in log-space. Industrial steady operation often includes feed/discharge and classification; these open-system effects can be incorporated as source/sink terms or boundary fluxes, but are outside the present scope. The numerical validation included here targets: (a) forward evolution accuracy and stability, (b) reduction accuracy of drift--diffusion, and (c) inverse identifiability under noise on a controlled benchmark.

\section{Conclusions}
We formulated pure fragmentation as conservative transport in log-size space, yielding an exact one-sided jump master equation for normalized mass fraction. We presented an optional GKSL/Lindblad factorization of the classical generator as a constrained parameterization for inverse modeling and a compact semigroup language, without claiming it is necessary for forward simulation. Under a small-jump regime we obtained a drift--diffusion reduction and a standard detailed-balance symmetrization that motivates explicit low-dimensional spectral hypothesis classes for PSD shapes. Finally, we provided numerical validation: forward jump vs.\ drift--diffusion comparisons and inverse parameter recovery under noise on a well-posed synthetic benchmark.

\appendix

\section{Appendix A: A convenient exponential-jump family (worked moments)}
A simple self-similar family is $B(z)=\beta z^{\beta-1}$ for $z\in(0,1)$, $\beta>0$, which implies $K(u)=\beta \ee^{-\beta u}$ and moments $m_1=1/\beta$, $m_2=2/\beta^2$.

\section{Appendix B: Non-self-adjoint caveat for open steady states}
\label{app:biorth}
In open fragmentation settings (feed in, product out, or nonzero size-space flux), reduced operators can be non-self-adjoint and left/right modes may be needed. This appendix records the caveat: in such cases steady densities need not be representable as a single modulus square; rather, left/right eigenmodes may be required in non-self-adjoint spectral reductions. (A full open-system comminution model with classification is outside the present scope.)


\end{document}